\documentclass[11pt]{article}
\usepackage{moriond,epsfig}

\def\Journal#1#2#3#4{{#1} {\bf #2}, #3 (#4)}

\def\PRL{\em Phys. Rev. Lett.}
\def\PRD{{\em Phys. Rev.} D}

\def\be{\begin{equation}}
\def\ee{\end{equation}}
\def\bea{\begin{eqnarray}}
\def\eea{\end{eqnarray}}

\begin{document}
\vspace*{4cm}
\title{Soft QCD at Tevatron}

\author{M. Rangel \\
 on behalf of the D0 and CDF collaborations}

\address{
Laboratoire de l'Acc\'el\'erateur Lin\'eaire, Universit\'e Paris-Sud 11, B\^atiment 200 \\
91898, Orsay, France \\
}

\maketitle\abstracts{
Experimental studies of soft Quantum Chromodynamics (QCD) at
Tevatron are reported in this note.
Results on inclusive inelastic interactions, underlying events, double
parton interaction and exclusive diffractive production
and their implications to the Large Hadron Collider (LHC) physics are discussed.
}

\section{Introduction}

In hadron collisions, hard interactions are theoretically defined as
collisions of two incoming partons along softer interactions from the
remaining partons. 
The soft effects become especially
important in very high luminosity enviromments (such as the
Large Hadron Collider) and they need to be accounted
for in most of experimental measurements. 
In a particular case, the incoming
hadrons stay intact after the collision producing a clean signature,
which can be used to search for new physics.

In this note, we review several experimental results related to soft
QCD and their implications to LHC physics: inclusive inelastic interactions 
(section \ref{inelastic}), 
underlying event (section \ref{ue}), double parton interactions 
(section \ref{doubleparton})
and exclusive diffractive production (section \ref{edp}).

\section{Inclusive inelastic interaction \label{inelastic} }

The so-called ``minimum-bias" (MB) interactions are defined
as data collected with a trigger set up so as events are selected
with uniform acceptance from all possible inelastic interactions.
Description of inelastic nondiffractive events can only be
accomplished by a nonperturbative phenomenological model such
as that made available by the PYTHIA Monte Carlo (MC) generator.

Different observables of the final state of antiproton-proton
interactions measured with the CDF detector were compared to
PYTHIA Tune A \cite{mb}. Both the charged and neutral particle
activites were studied. In general, poorly agreement is observed
between existing MC and data, and the measurements can be used
to improve QCD MC.

\begin{figure}
\begin{center}
\psfig{figure=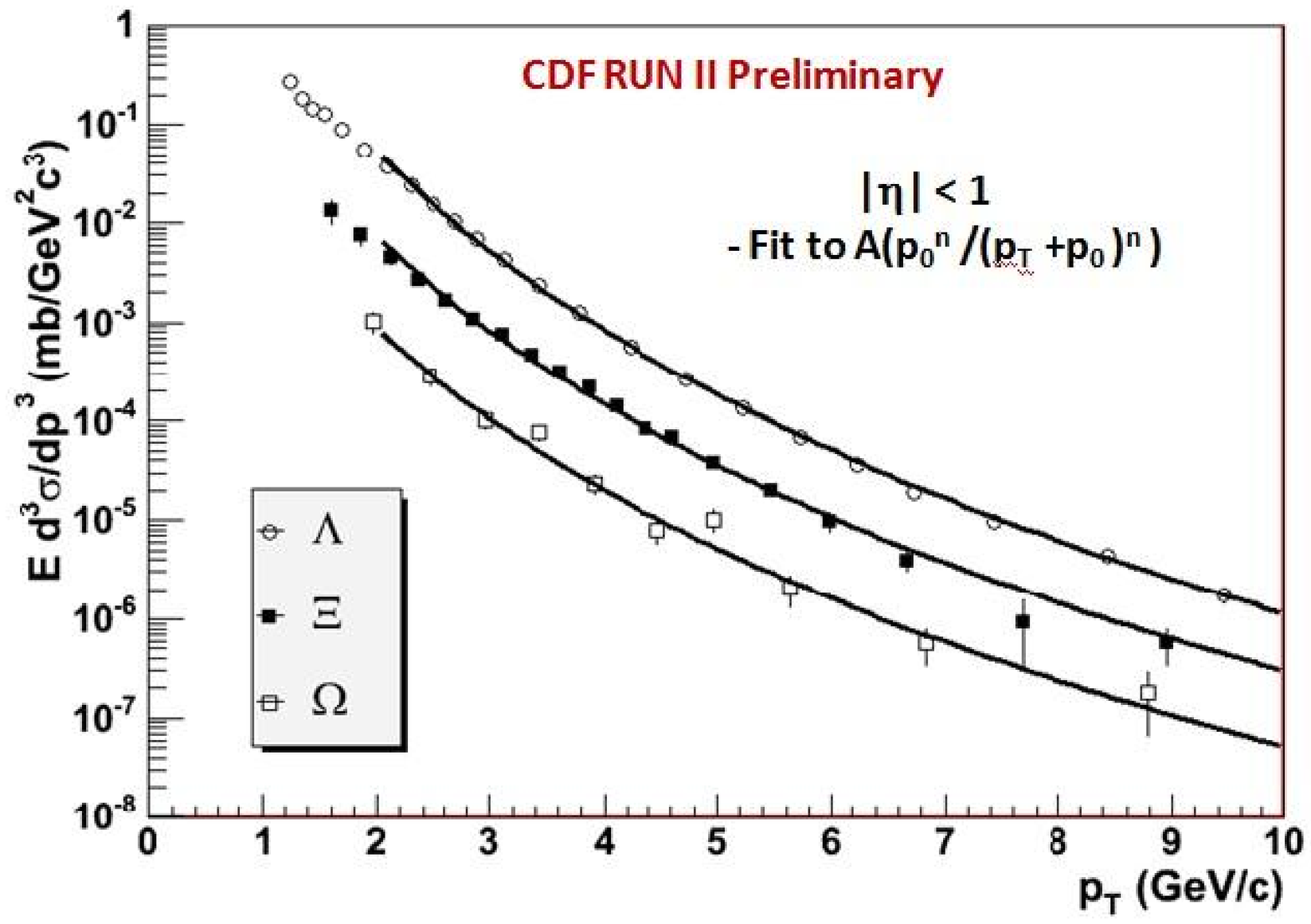,height=2.0in}
\psfig{figure=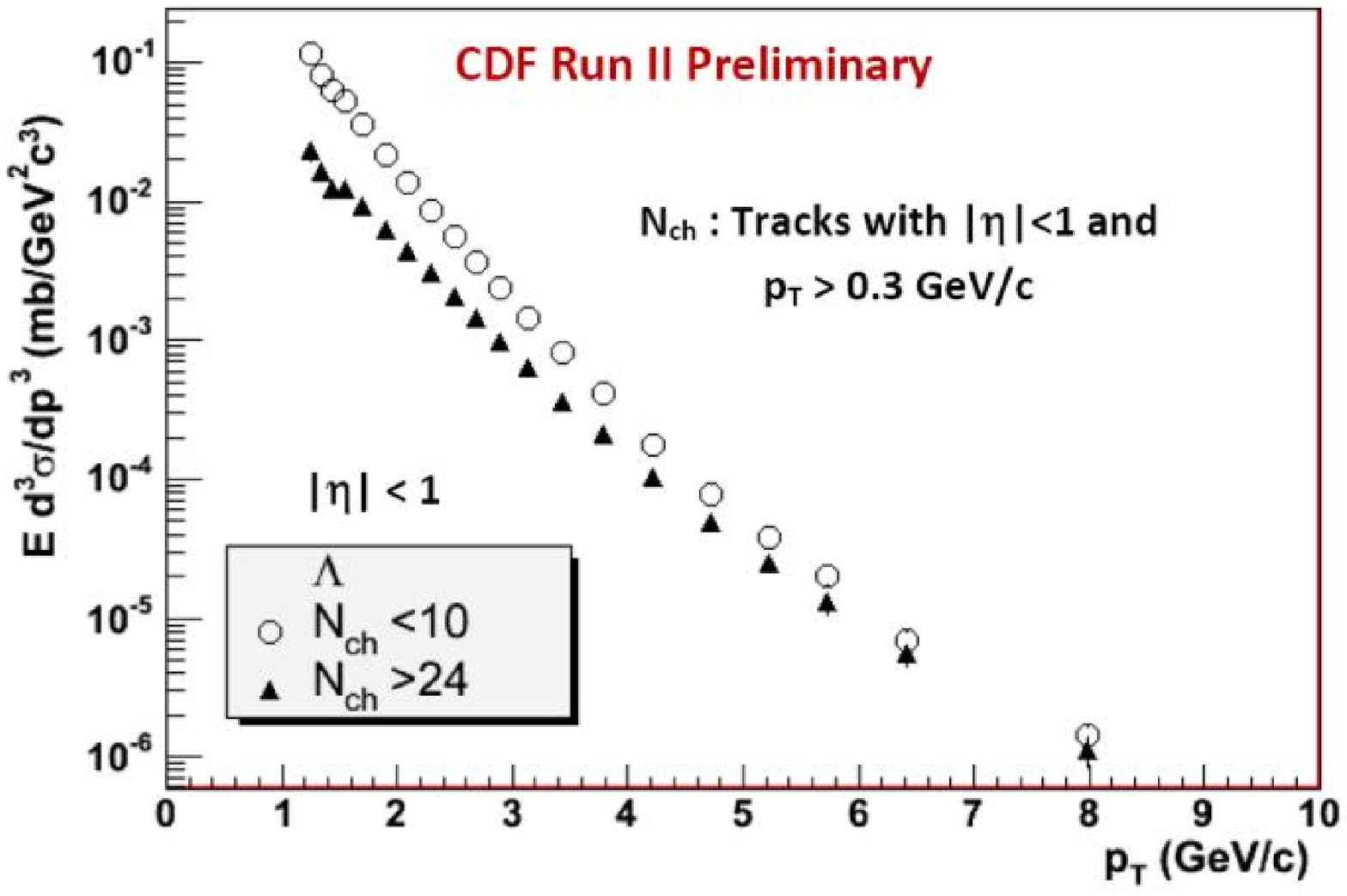,height=1.7in}
\end{center}
\caption{
Left upper figure: Inclusive invariant $p_T$ distribution for $\Lambda$, $\Sigma$ and $\Omega$ within $|\eta| <$1.
The solid curves are from fit to the functional form $(A)(p0)^n /(pT +p0)^n$ with $p0$ =1.3. 
Right upper figure: The ratio of $\Sigma / \Lambda$ and $\Omega / \Lambda$ as a function of $p_T$.
Two lower figures: The inclusive pT distributions for two different multiplicity regions,
number of charged particles $<$ 10 and $>$ 24. Left plot is for
lambdas and right plot is for cascades.
\label{fig:hyp}}
\end{figure}

Measurements of inclusive invariant $p_T$ differential cross section
of centrally produced hyperons ($|\eta| <$ 1) were performed in
minimum biased events \cite{mbhyp}. Cascades ($\Sigma$), omegas ($\Omega$)
and lambdas ($\Lambda$) particles are selected and their $p_T$
spectrum measured (Fig. \ref{fig:hyp}). It is observed that
the production ratio of the three particles is fairly constant as a function of $p_T$.

\section{Underlying Event \label{ue}}

The underlying event (UE) consists of the beam-beam remnants minus the hard-scattering
products and is becoming increasingly important to the discovery and precision potential at hadron
colliders. CDF has conducted UE studies that exploit jet and Drell-Yan
event activity topologies to maximize the sensitivity of UE observables \cite{ue}.
Several distributions of UE-sensitive observables, corrected to the particle level, suggest the
UE may be universal (independent of the hard process) and inform MC tuning and development.

A good agreement between data and PYTHIA Tune AW Monte Carlo predictions was
observed, except by a slight excess at transverse region compared to toward region,
which is caused by transverse regions receiving contributions from away side jet \cite{ue}.

\section{Double parton interaction \label{doubleparton}}

D0 has studied $\gamma$+3-jet events to measure double parton scattering (DPS), whereby two pairs
of partons undergo hard interactions in a single $p\bar{p}$ collision. DPS is not only a background to
many rare processes, especially at higher luminosities, but also provides insight into the spatial
distribution of partons in the colliding hadrons.
The DPS cross section is expressed as $\sigma_{{DPS}^{\gamma+3jet}}$ = $\sigma_{jj}\sigma_{\gamma j}/\sigma_{eff}$,
where $\sigma_{eff}$ is the effective interaction region that decreases for less uniform spatial
parton distributions. D0 measures a mean of $\sigma_{eff}$ = 16.4 $\pm$ 2.3 mb \cite{dpd0}, which is consistent with an
earlier CDF result \cite{dpcdf}, and finds $\sigma_{eff}$ to be independent of jet $p_T$ in the second interaction. 
More precise studies can reveal a $\sigma_{eff}$ sensitivity to jet $p_T$, 
this could indicate a dynamical departure from 
the na\"ive assumption that DPS depends on an uncorrelated product of
$\sigma_{jj}$ and $\sigma_{\gamma j}$.

\section{Exclusive Diffractive Production \label{edp}}

Exclusive diffrative processes are those where the colliding hadrons emerge intact, but part of their
momentum is lost producing central objetcs, with surrounding rapidity regions devoid of particles.
CDF recently reported observations of $p\bar{p} \rightarrow p$+dijet+$\bar{p}$, with $p_T^{jet} >$ 10 GeV;
and $p\bar{p} \rightarrow p$+[$\mu^{-}\mu^{+},J/\Psi,\Psi(2S),\chi_C^{0}$]+$\bar{p}$
with two oppositely charged central muons and either no other particles or one additional photon detected \cite{edpcdf}.

\begin{figure} [t]
\begin{center}
\psfig{figure=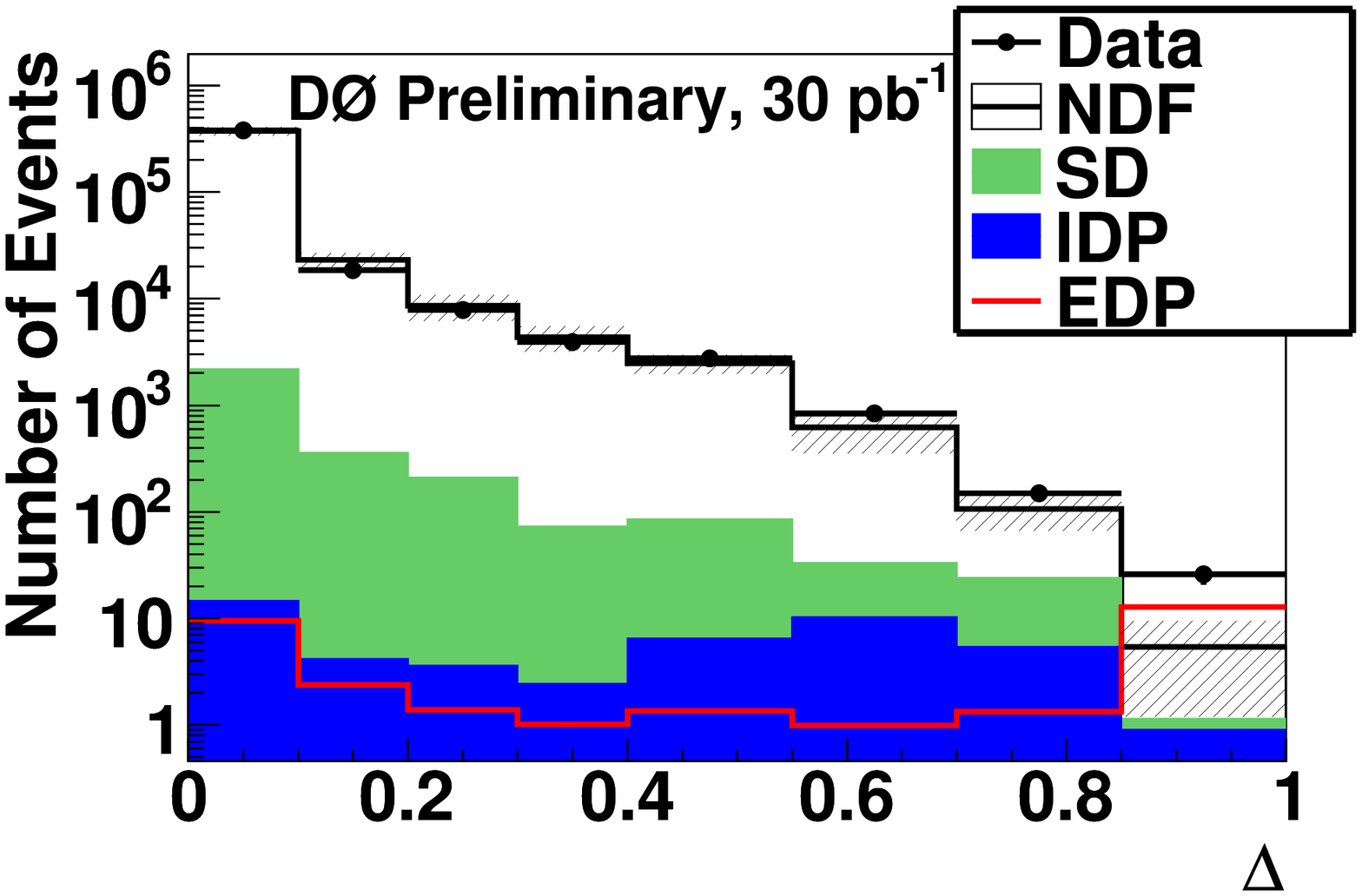,height=1.7in}
\psfig{figure=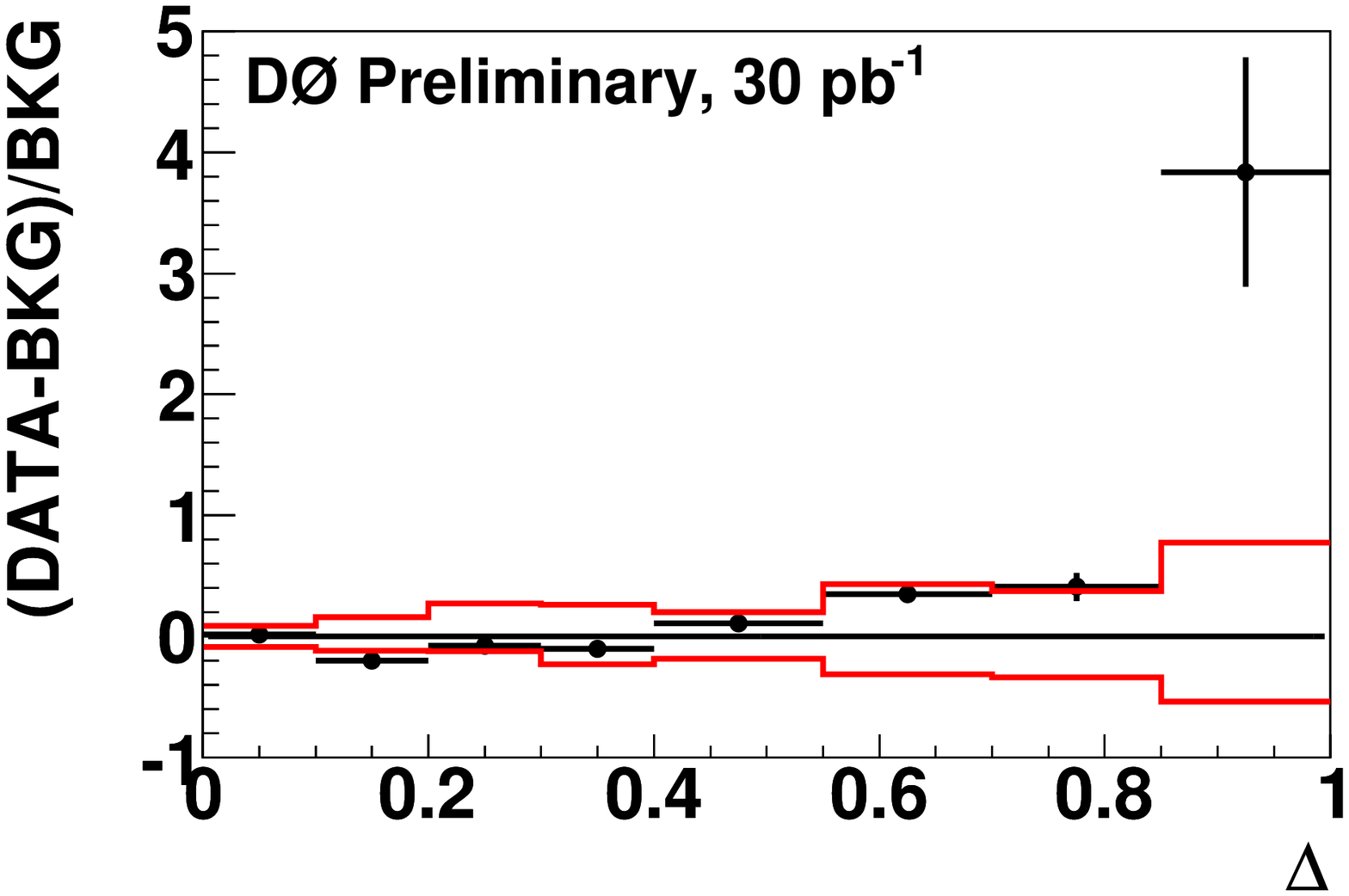,height=1.7in}
\psfig{figure=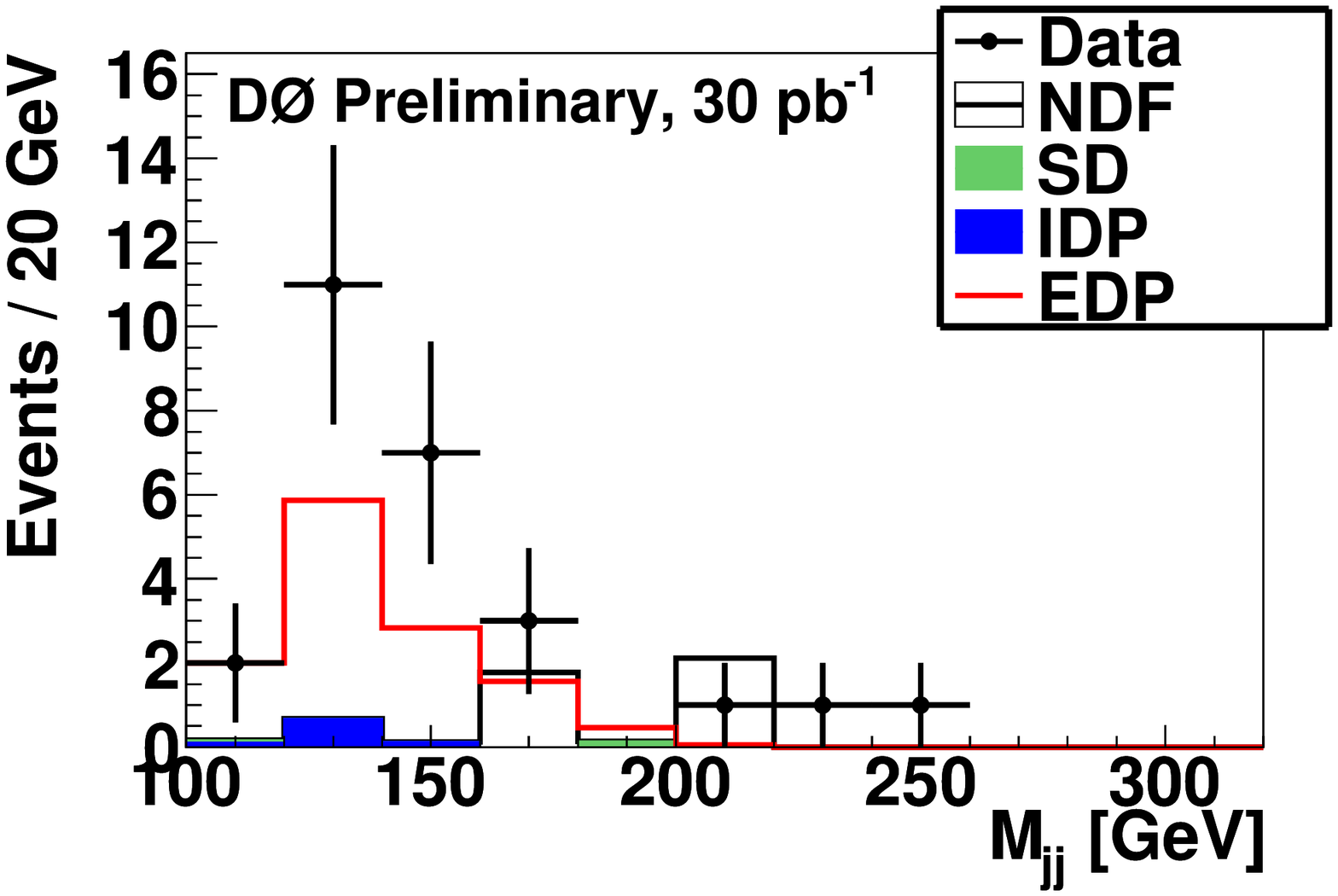,height=1.7in}
\psfig{figure=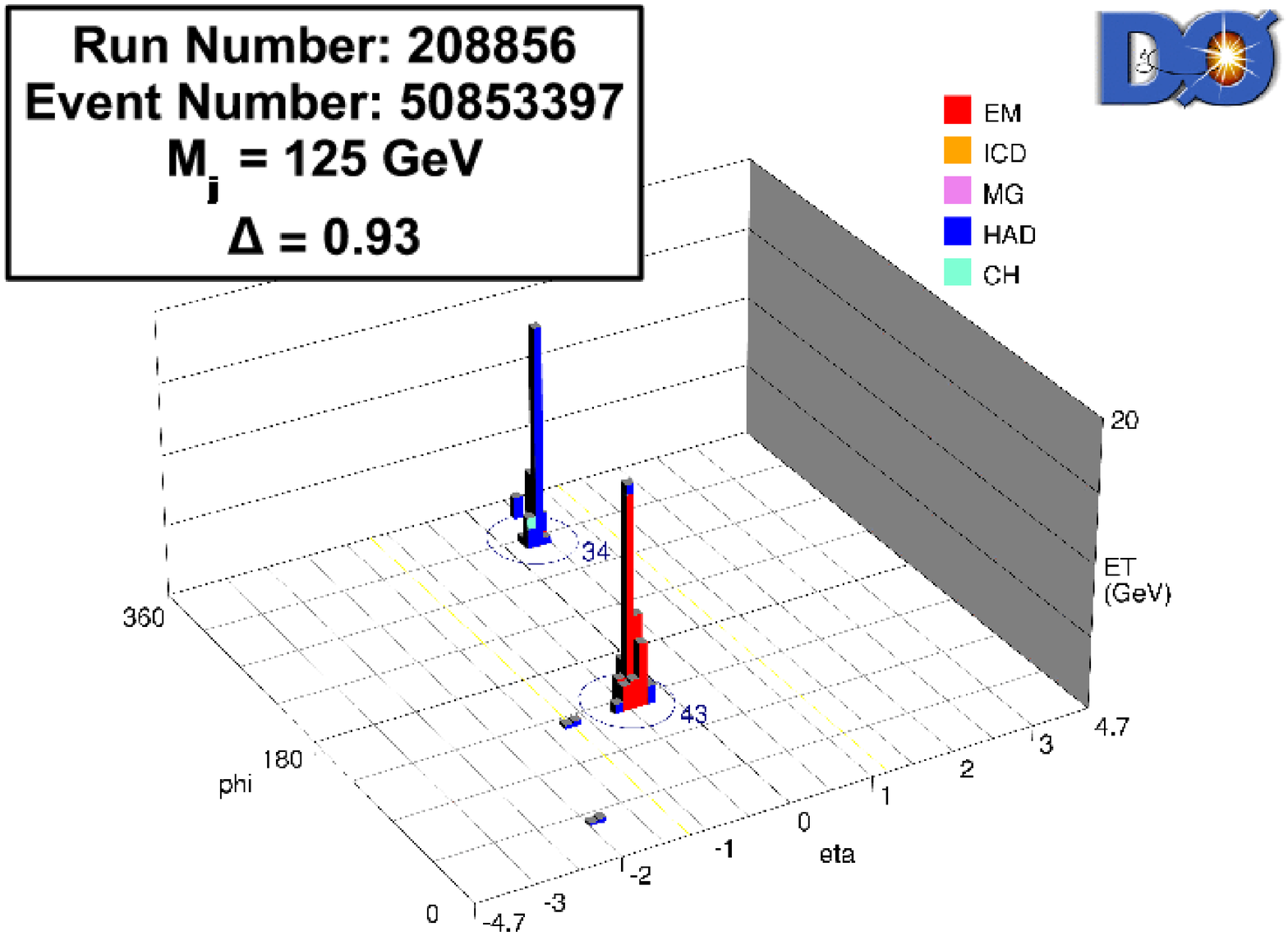,height=1.7in}
\end{center}
\caption{
Top Left Figure: $\Delta$ distribution for data and background 
(nondiffractive (NDF), single-diffractive (SD) and inclusive double pomeron (IDP)).
A good agreement is observed between data and background except
at high values of $\Delta$ where exclusive diffracitve production (EDP) dominates. 
The hatched band indicates the total uncertainty on the background. 
Top Right Figure: MC background (BKG) subtracted data divided by background. 
The solid lines are $\pm$ 1 standard deviation systematic uncertainty
on the background.
Bottom Left Figure: Dijet invariant mass distribution for MC and data after applying
the cut on $\Delta \geq$ 0.85. The total background prediction
is of 5.4 $^{+4.2}_{-2.9}$ events and 26 signal candidate events are observed in data.
Bottom Right Figure: Exclusive diffractive candidate event. No energy deposition is
present in the forward regions, only two central jets are observed in the detector
\label{fig:edp}}
\end{figure}

D0 recently reported an evidence for diffractive exclusive dijet production with an invariant mass greater than
100 GeV. A discriminant variable ($\Delta$ based on calorimeter information was used to demonstrate 
a significant excess of events with very little energy outside the dijet system (Fig. \ref{fig:edp}). 
The probability for the observed excess to be explained by
other dijet production processes is 2 $\times$ 10$^{−5}$ , corresponding to a 4.1 standard deviation significance \cite{edpd0}.

Tevatron results support the viability of exclusive Standard Model Higgs production through $p$+H+$p$ 
processes at the LHC, which are expected to play an important role in future studies of new physics
\cite{kmr}.

\section{Conclusion}

The Tevatron experiments provide soft QCD physics studies at $\sqrt{s}$ = 1.96 TeV.
These results are based on less than one third of the anticipated complete Run II sample,
therefore more measurements are expected in the coming years to illuminate the 
Large Hadron Collider physics results.


\section*{References}
\begin{thebibliography}{99}

\bibitem{mb}
  T.~Aaltonen {\it et al.}  [CDF Collaboration], \Journal{\PRL}{79}{112005}{2009}.
\bibitem{mbhyp}
  CDF Collaboration, CDF Public Note 10084.
\\ Web Page: http://www-cdf.fnal.gov/physics/new/qcd/hyperons\_10/HyperonWEBV2.htm.
\bibitem{ue}
  CDF Collaboration, CDF Public Note 9351. \\
  Web Page: http://www-cdf.fnal.gov/physics/new/qcd/run2/ue/chgjet/index.html.

\bibitem{dpd0}
  V.~M.~Abazov {\it et al.}  [D0 Collaboration],  \Journal{\PRD}{81}{052012}{2010}.

\bibitem{dpcdf}
  F. Abe {\it et al.}  [CDF Collaboration],  \Journal{\PRD}{56}{3811}{1997}.

\bibitem{edpcdf}
  A.~Abulencia {\it et al.}  [CDF Collaboration], \Journal{\PRL}{98}{112001}{2007}.
\\  T.~Aaltonen {\it et al.}  [CDF Collaboration], \Journal{\PRL}{99}{242002}{2007}.
\\  T.~Aaltonen {\it et al.}  [CDF Collaboration], \Journal{\PRD}{77}{052004}{2008}.
\\  T.~Aaltonen {\it et al.}  [CDF Collaboration], \Journal{\PRL}{102}{222002}{2009}.
\\  T.~Aaltonen {\it et al.}  [CDF Collaboration], \Journal{\PRL}{102}{242001}{2009}.
\bibitem{edpd0}
  D0 Collaboration, D\O\ Note 6042-CONF. \\
  Web Page: http://www-d0.fnal.gov/Run2Physics/WWW/results/prelim/QCD/Q17/.

\bibitem{kmr}
  V.~A.~Khoze, A.~D.~Martin and M.~G.~Ryskin, Eur.\ Phys.\ J.\  C {\bf 23}, 311 (2002).
\end{thebibliography}
\end{document}